\def\Roman#1{\uppercase\expandafter{\romannumeral#1}}

\documentclass[12pt]{article}

\usepackage[cp1251]{inputenc}
\usepackage[english]{babel}
\usepackage{amssymb,amsfonts}

\begin{document}


\author{S. N. Storchak \\
\small{Institute for High Energy Physics,
Protvino,}\\
\small{Moscow Region, 142284, Russia,}\\
\small{e-mail: storchak@mx.ihep.su}}

\title{Dependent coordinates in path integral measure 
factorization}

\date{ }
\maketitle

\begin{abstract}
The transformation of the path integral measure under the reduction
procedure in the dynamical systems with a symmetry is considered. The
investigation is carried out in the case of the Wiener--type path
integrals that are used for  description of the diffusion on a smooth
compact Riemannian manifold with the given free isometric action of
the compact semisimple unimodular Lie group. The transformation of the
path integral, which  factorizes the path integral measure, is based
on the application of the optimal nonlinear filtering equation from
the stochastic theory.
The integral relation between the kernels of the original and reduced
semigroup are obtained.\\


\end{abstract}

	\section{Introduction}
The  standard approach to the path integral quantization of the gauge
field theories is based on the Faddeev---Popov method \cite{Faddeev},
by which a path integral over  invariant variables is rewritten  as a
path integral over  variables constrained by some gauge conditions.
But in order to obtain  such a representation  it is necessary  to
separate the path integral measure on two parts related,
correspondingly,  with the gauge--invariant (independent) and the
gauge--dependent (or group) degrees of freedom.   
However, in general it is unknown: does this separation of the path
integral measure  leads (or not) to  a some Jacobian in the path
integral measure.

In the present paper we consider the path integral quantization
problem for the scalar particle which moves on a manifold with a given
group action. 
 In this problem, as in gauge field theories,  an original manifold
 can be regarded  as a total space of the principle fiber bundle. As a
 consequence of such a  representation, we get a new way
   of the coordinatization of an initial  manifold. The coordinate
   functions of a chart  in the manifold atlas  can be  given by
   invariant variables related with the orbit space and by the
   group--valued variables defined by the group element which
   "measures"  the distance  between the considered point in the total
   space of the fiber bundle and the base point which can be  reached
   along the orbit.

  The path integral quantization of our problem was considered in
  \cite{Tanimura,Kunstatter} by using the definitions of the path
  integrals based on discrete approximations.
In our papers \cite{Stor1,Stor2}, we studied the quantization of this
problem with the help of the path integrals in which the path integral
measures were defined by the stochastic processes. It was found there,
that the factorization of the path integral measure can be performed
by applying the nonlinear filtering equation from the stochastic
process theory.

The description of the orbit space evolution in gauge field theories
are usually given  by the constrained (or dependent) coordinates. It
means, that the corresponding  variables should meet the additional
conditions 
(the gauges). 
From the local point of view, these additional conditions (given by
the system of  equations) define a local submanifold in the original
manifold. In turns, the submanifold can be regarded as a local section
of the principal fiber bundle. 

Since there is a local isomorphism between the initial principle fiber
bundle and the trivial principal fiber bundle which has  
 this local submanifold  as the base space \cite{Mitter}, 
we can also make use   these dependent coordinates as the coordinates
on our principal fiber bundle.

If these local submanifolds (the local sections) are the parts of a
some global submanifold (a global section) then our initial principal
fiber bundle is a trivial one. 
And in this case, our dependent coordinates have the global meaning.
But in general, there is no a global section in the principal fiber
bundle.	So, in a local neighborhood of the each point of the initial
manifold, one should  introduce  its own  dependent coordinates. 

 After performing the local factorization of the path integral measure
 into the "the group measure" and the measure that are given on the
 local section, one should to solve 
  the problem of the definition of a some global measure related with
  the set of these local measures. 
  
  Recently \cite{Kelnhofer}, it was found the possible solution of
  such a problem. In this paper it  was shown    a way of description
  of the global path integral measure
 in terms of   the local measures.
    Therefore, from the principal point of view,  the problem
of the introduction of the dependent coordinates leads us to the
consideration of the particular case of the trivial principal fiber
bundle. 

The representation of the orbit space path integrals as the integral
over the dependent  variables were studied in many papers (see, for
example, \cite{Jaskolski,Falck,Kunstatter1}). But the path integral
measure factorization
questions  was not considered there.
In the present   paper, we will investigate the behavior of the path
integral measure under changing the path integral variables for the
dependent ones in the Wiener--like path integrals that are used to
describe the "quantum" motion of the scalar particle on a smooth
compact Riemannian manifold (without boundary) on which the free
effective isometric action of the compact semisimple unimodular Lie
group  is given. 

\section{Definitions}
The objects of our consideration will be the path integrals
representing the solution of the backward Kolmogorov equation given on
a smooth compact Riemannian manifold $\cal P$:
\begin{equation}
\left\{
\begin{array}{l}
\displaystyle
\left(
\frac \partial {\partial t_a}
+\frac 12\mu ^2\kappa \triangle
_{\cal P}(p_a)+\frac
1{\mu ^2\kappa m}
V(p_a)\right){\psi}_{t_b} (p_a,t_a)=0\\
{\psi}_{t_b} (p_b,t_b)=\phi _0(p_b),
\qquad\qquad\qquad\qquad\qquad (t_{b}>t_{a}),
\end{array}\right.
\label{1}
\end{equation}

In this equation $\mu ^2=\frac \hbar m$ ,
$\kappa $ is a real positive parameter,
$\triangle _{\cal P}(p_a)$  is a Laplace--Beltrami operator on
manifold $\cal P$, and $V(p)$ is a group--invariant potential term. In
a chart $({\cal U},{\varphi}^A )$ with the coordinate functions $Q^A=
{\varphi}^A (p)$,  the Laplace -- Beltrami operator is given as 
\[
\triangle _{\cal P}(Q)=G^{-1/2}(Q)\frac \partial
{\partial Q^A}G^{AB}(Q)
G^{1/2}(Q)\frac\partial {\partial Q^B},
\]
where the matrix $G^{AB}(Q)$ is inverse to the matrix  of the
Riemannian metric $G_{AB}$ given in the the coordinate basis
$\{\frac{\partial}{\partial Q^A}\}$, $G=det (G_{AB})$.

In order to get the Schr\"odinger equation from the equation (\ref{1})
one should perform the transition to the forward Kolmogorov equation
and then to put  $\kappa =i$ in obtaining equation. In case of some
analytical restrictions imposed on the coefficients of the  the
backward Kolmogorov equation, its fundamental solution satisfies also
a forward equation. However, the transition from the Wiener--like path
integrals to the Feynman' ones is a special problem which needs an
additional investigation and does not considered in the paper. 

There are different representations of
the solution of the equation (\ref{1})  in terms of the path integral.
We will use the definition of the path integral from \cite{Dalecky},
together with the assumptions that all necessary analytical conditions
for this  take place in our paper. 
By \cite{Dalecky}, the solution of (\ref{1}) can be written as
follows:
\begin{eqnarray}
{\psi}_{t_b} (p_a,t_a)&=&{\rm E}\Bigl[\phi _0(\eta (t_b))
\exp \{\frac 1{\mu
^2\kappa m}\int_{t_a}^{t_b}V(\eta (u))du\}\Bigr]\nonumber\\
&=&\int_{\Omega _{-}}d\mu ^\eta (\omega )
\phi _0(\eta (t_b))\exp \{\ldots \},
\label{2}
\end{eqnarray}
where  ${\eta}(t)$ is a global stochastic process on a manifold $\cal
P$, 
${\mu}^{\eta}$ is the path integral measure on the path space $\Omega
_{-}=\{\omega (t):\omega (t_a)=0,
\eta (t)=p_a+\omega (t)\}$   defined by the probability distribution
of a stochastic process ${\eta}(t)$.

On charts $({\cal U},\varphi)$ in the atlas for the manifold $\cal P$,
the global stochastic process $\eta (t)$ is defined by the local
processes $\varphi (\eta)=\eta _{\varphi }(t)\equiv
\{\eta ^{A}(t)\}$ that are    solutions of  the stochastic
differential equations:
\begin{equation}
d\eta ^A(t)=\frac12\mu ^2\kappa G^{-1/2}\frac \partial {\partial
Q^B}(G^{1/2}G^{AB})dt+\mu \sqrt{\kappa }{\mathfrak X}_{\bar{M}}^A(\eta
(t))
dw^{
\bar{M}}(t),
\label{3}
\end{equation}
where the matrix ${\mathfrak X}_{\bar{M}}^A$ is defined 
by the local equality  
$\sum^{n_{P}}_{\bar{
{\scriptscriptstyle K}}
\scriptscriptstyle =1}
{\mathfrak X}_{\bar{K}}^A{\mathfrak X}_
{\bar{K}}^B=G^{AB}$.\\
 Here and in what follows  we denote the Euclidean indices by
 over--barred indices. 

According to \cite{Dalecky}, the global semigroup determined by  the
equation (\ref{2}) acts in the space of  smooth and bounded functions
on $\cal P$. It is defined by the limit (under the refinement of the
time interval) of the superposition of the local semigroups
\begin{equation}
\psi _{t_b}(p_a,t_a)=U(t_b,t_a)\phi _0(p_a)=
{\lim}_q {\tilde U}_{\eta}(t_a,t_1)\cdot\ldots\cdot
{\tilde U}_{\eta}(t_{n-1},t_b)
\phi _0(p_a),
\label{5}
\end{equation}
where, in turns, each of the local semigroup  
${\tilde U}_{\eta}$ is related with the local representative of the
global stochastic process $\eta $.

One of the main advantage of the definition (\ref{5}) is that it 
 gives us an opportunity to derive the transformation properties of
 the path integral of (\ref{2}) by studying  the local semigroups
 ${\tilde U}_{\eta}$\footnote{In the following, we omit the the
 potential term of the Hamiltonian operator as it is inessential for
 us in performing the path integral transformations. It will be
 recovered in  final formulas.}: 
 \[
 {\tilde U}_{\eta}(s,t) \phi (p)={\rm E}_{s,p}\phi (\eta (t)),\,\,\,
 s\leq t,\,\,\,\eta (s)=p,
 \]

 These local semigroup are also defined by the path integrals with the
 integration measures determined by the local representatives $\varphi
 ^{\cal P}(\eta (t))={\eta}^{\varphi
^{\cal P}}(t)\equiv\{\eta ^A (t)\}$ of the global stochastic process
$\eta (t)$.

\section{Principal fiber bundle coordinates}
The problem, which we consider in the paper, is related with the
investigation of the reduction procedure in the dynamical systems with
the symmetry.
Due to the symmetry, the initial dynamical system is reduced to the
system that can be described in terms of the invariant variables.
The geometry of this problem is well developed \cite{AbrMarsd}.

A free (effective) action of the compact semisimple group Lie $\cal G$
on a smooth compact manifold $\cal P$\footnote{In our case this is an
isometric action on a Riemannian manifold.}
 leads to the orbit--fibering of the manifold $\cal P$.
And we can regard the manifold $\cal P$ as a total space of the
principal fiber bundle $P(\cal M,\cal G)$ where the orbit space 
$\cal M$ is a base space.

The principal bundle picture means that locally manifold  
$\cal P$ has the product structure: 
${\pi}^{-1}({\cal U}_{x})\sim {\cal U}_{x}\times \cal G$,
where ${\cal U}_{x}$ is an open neighborhood of the point $x=\pi (p)$
which belongs to the chart 
$({\cal U}_x,\varphi _x)$ of the bundle $P(\cal M,\cal G)$. 
Therefore,  we can equally use the principal bundle coordinates for
the   coordinatization of our manifold $\cal P$.
In other words, we can express an initial coordinates $Q^A$ of the
point $p$ in terms of the principal bundle coordinates 
$(x^i,a^{\alpha})$ ($i=1,\dots,N_{\cal M}$,
   $N_{\cal M}=dim {\cal M}$, $\alpha =1,
   \dots,N_{\cal G}$, $N_{\cal G}=dim {\cal G}$, 
$N_{\cal P}=N_{\cal M}+N_{\cal G}$).

We could take the set of the functionally independent and $\cal
G$--invariant functions that are  solutions of the special
differential equations as  the invariant coordinates $x^i(Q)$ (the
orbit space coordinates) of a point.
However, in many cases  finding of solutions of these differential
equations becomes a very difficult problem.
Hopefully, there is also an  another method of the orbit space
coordinatization in which the necessary invariant coordinates are
introduced with the help of the gauge constraints. 

It is supposed, that in each sufficiently small neighborhood of the
point $p$, belonging to the manifold   $\cal P$, there are  a set of
the functions $\{\chi ^{\alpha}(Q),\,\,\, \alpha=1,\dots,
 N_{\cal G}\}$, that can be used (by the equations
 ${\chi}^{\alpha}(Q)=0$) to determine a some local submanifold of the
 manifold $\cal P$. It is required that the submanifold  should have
 the transversal intersection with each of the orbits.
Then the coordinates on the manifold $\cal P$ can be introduced as
follows.

By our assumption, we have the action of the group 
$\cal G$ on the manifold $\cal P$:  $\tilde p=pa$,  or in coordinates: 
$\tilde Q^A= F^A(Q^B,a^{\alpha})$, 
where $Q^A$ are the coordinates of a point $p$. We assume that it is a
right action, i.e., 
$(pa_1)a_2=p(a_1a_2)$:
\[
F(F(Q,a_1),a_2)=F(Q,\Phi(a_1,a_2))
\]
where $\Phi$ is the function which determines the group multiplication
law in the space of the group parameters.

The group coordinates $a^{\alpha}(Q)$  of a point $p$ are defined as a
coordinates of  that  group element which  carries  the point $p$ to
the local submanifold $\{{\chi}^{\alpha}(Q)=0\}$.  These group
coordinates are given by the solution of the following equation:
\[
{\chi}^{\alpha}\bigl(F^A(Q,a^{-1}(Q))\bigr)=0.
\]

The invariant coordinates $x^i(Q)$ of $p$ are the coordinates of that
point of the submanifold $\{{\chi}^{\alpha}(Q)=0\}$ which is obtained
from the point $p$ under the action of the group element with the
coordinates $a^{\alpha}(Q)$.
 If the submanifold $\{{\chi}^{\alpha}(Q)=0\}$ is given
 parametrically: $Q^A=Q^{\ast}{}^A(x^i)$, 
the coordinates $x^i(Q)$ are defined by the equation:
\[
Q^{\ast}{}^A(x^i)=F^A(Q,a^{-1}(Q)).
\]
We refer to \cite{Razumov} where this way of the coordinatization was
considered and where the geometrical generalization of the Bogolubov
coordinate transformation method from \cite{Khrustalev} was obtained.

The path integral transformation induced by replacement of coordinates
$Q^A$ for $(x^i(Q), a^{\alpha}(Q))$  and the factorization of the path
integral measure was considered in \cite{Stor2}. In the present paper
we study the same problem, but now together with the group coordinates
$a^{\alpha}$ (obtained by previous method)
we will use   the constrained (or dependent) coordinates 
$Q^{\ast}{}^A\,\,$:
 $\{{\chi}^{\alpha}(Q^{\ast}{}^A)=0\}$ 
of the corresponding point  on the local submanifold
$\{{\chi}^{\alpha}=0\}$
 of the original manifold $\cal P$.
 
We will assume, that these local submanifolds form a global
submanifold $\Sigma$ in the original manifold $\cal P$. Therefore, our
principal fiber bundle $P(\cal M,\cal G)$ is a trivial one. 

Provided that the coordinates $Q^{\ast}$ are constrained:
$\{{\chi}^{\alpha}(Q^{\ast})=0\}$, the initial coordinates
${Q^A}$ of a point $p$ are given by the equation
$Q^A=F^A(Q^{\ast}{}^A,a^{\alpha})$. 
Later we will see that an apparent ambiguity of a  transition from
$Q^A$ to $(Q^{\ast}{}^A,a^{\alpha})$ is compensated by the presence of
the corresponding projection  operators in resulting expressions.

The representation of a Riemannian metric of the manifold $\cal P$ in
new coordinates $(Q^{\ast}{}^A,a^{\alpha})$ is derived from the
transformation of the coordinate vector fields.
It is given as follows:
\begin{eqnarray}
\frac{\partial}{\partial Q^{B}}&=&F^{C}_{B}
(F(Q^{\ast},a),a^{-1})N^{A}_{C}(Q^{\ast})
\frac{\partial}{\partial Q^{\ast}{}^A}\nonumber\\
&&+F^{E}_{B}(F(Q^{\ast},a),a^{-1}){\chi}^{\mu}_{E}(Q^{\ast})
(\Phi ^{-1}){}^{\beta}_{\mu}(Q^{\ast}){\bar v}^
{\alpha}_{\beta}(a)\frac{\partial}{\partial a^{\alpha}}.
\label{7}
\end{eqnarray}
Here $F^{C}_{B}(Q,a)\equiv \frac{\partial F^{C}}
{\partial Q^{B}}(Q,a)$, ${\chi}^{\mu}_{E}\equiv
\frac{\partial {\chi}^{\mu}}{\partial Q^{E}}(Q)$,
$(\Phi ^{-1}){}^{\beta}_{\mu}(Q)$ -- the matrix which is inverse to
the Faddeev -- Popov matrix:
\[
(\Phi ){}^{\beta}_{\mu}(Q)=K^{A}_{\mu}(Q)
\frac{\partial {\chi}^{\beta}(Q)}{\partial Q^{A}}
\]
($K_{\mu}$ are the Killing vector fields for the Riemannian metric 
$G_{AB}(Q)$), the matrix  ${\bar v}^{\alpha}_{\beta}(a)$ is inverse to
the matrix ${\bar u}^{\alpha}_{\beta}(a)$. The $\det {\bar
u}^{\alpha}_{\beta}(a)$ is a density of a right invariant measure
given on the group $\cal G$.

Finally, $N^{A}_{C}$ is a matrix form of the projection operator
($N^{A}_{B}N^{B}_{C}=N^{A}_{C}$) which project onto the   orthogonal
to the Killing vector field subspace:
\[
N^{A}_{C}(Q)={\delta}^{A}_{C}-K^{A}_{\alpha }(Q)
(\Phi ^{-1}){}^{\alpha}_{\mu}(Q){\chi}^{\mu}_{C}(Q).
\]
In (\ref{7}), this projection operator  is restricted to the
submanifold $\{{\chi}^{\alpha}=0\}$:
\begin{eqnarray*}
&&N^{M}_{D}(Q^{\ast})\equiv N^{M}_{D}(F(Q^{\ast},e)),\nonumber\\
&&N^{M}_{D}(Q^{\ast})=F^{B}_{D}(Q^{\ast},a)
N^{A}_{B}(F(Q^{\ast},a))
F^{M}_{A}(F(Q^{\ast},a),a^{-1}).
\end{eqnarray*} 
The formula (\ref{7}) is similar to the corresponding formula from
\cite{Falck,Creutz}.
On  treatment of the dependent coordinate we refer, for example, to
\cite{Plyushchay}. 

As an operator, the vector field $\frac{\partial}{\partial
Q^{\ast}{}^{A}}$  is defined by the rule:
\[
\left.
\frac{\partial}{\partial Q^{\ast}{}^{A}}
\varphi(Q^{\ast})=(P_{\perp})^{D}_{A}(Q^{\ast})
\frac{\partial \varphi (Q)}{\partial Q^{D}}
\right |_{Q=Q^{\ast}},
\]
where $P_{\perp}$ is a projection operator on the tangent plane to the
submanifold given by the gauges:
\[
(P_{\perp})^{A}_{B}=\delta ^{A}_{B}-{\chi}^{\alpha}_{B}
(\chi \chi ^{\top})^{-1}{}^{\beta}_{\alpha}(\chi ^
{\top})^{A}_{\beta}.
\]
Here $(\chi ^{\top})^{A}_{\beta}$ is a transposed matrix to the matrix
$\chi ^{\nu}_{B}$:
\[
(\chi ^{\top})^{A}_{\mu}=G^{AB}{\gamma}_
{\mu \nu}\chi ^{\nu}_{B},\,\,\, {\gamma}_{\mu \nu}
=K^{A}_{\mu}G_{AB}K^{B}_{\nu}.
\]
The above projection operators have the following properties:
\[
(P_{\perp})^{\tilde A}_{B}N^{C}_{\tilde A}=
(P_{\perp})^{C}_{B},\,\,\,\,\,\,\,\,\,N^{\tilde A}_
{B}(P_{\perp})^{C}_{\tilde A}=N^{C}_{B}.
\]

In new coordinate basis the metric $G_{AB}$ is written as a
metric ${\tilde G}_{\cal A\cal B}(Q^{\ast},a)$ with the following
components:
\begin{equation}
\left(
\begin{array}{cc}
G_{CD}(Q^{\ast})(P_{\perp})^{C}_{A}
(P_{\perp})^{D}_{B} & G_{CD}(Q^{\ast})(P_{\perp})^
{D}_{A}K^{C}_{\mu}\bar{u}^{\mu}_{\alpha}(a) \\
G_{CD}(Q^{\ast})(P_{\perp})^
{C}_{A}K^{D}_{\nu}\bar{u}^{\nu}_{\beta}(a) & {\gamma }_{\mu \nu }
(Q^{\ast})\bar{u}_\alpha ^\mu (a)\bar{u}_\beta ^\nu (a)
\end{array}
\right),
\label{8}
\end{equation}
where the projection operators $P_{\perp}$ depend on $Q^{\ast}$, i.e.,
they are restricted to the submanifold, 
$G_{CD}(Q^{\ast})\equiv G_{CD}(F(Q^{\ast},e))$:
\[
G_{CD}(Q^{\ast})=F^{M}_C(Q^{\ast},a)F^N_D(Q^{\ast},a)
G_{MN}(F(Q^{\ast},a)).
\]

The pseudoinverse matrix ${\tilde G}^{\cal A\cal B}(Q^{\ast},a)$ to
the matrix (\ref{8}) is as follows: 
\begin{equation}
\displaystyle
\left(
\begin{array}{cc}
G^{EF}N^{C}_{E}
N^{D}_{F} & G^{SD}N^C_S{\chi}^{\mu}_D
(\Phi ^{-1})^{\nu}_{\mu}{\bar v}^{\sigma}_{\nu} \\
G^{CB}{\chi}^{\gamma}_C (\Phi ^{-1})^{\beta}_{\gamma}N^D_B
{\bar v}^{\alpha}_{\beta} & G^{CB}
{\chi}^{\gamma}_C (\Phi ^{-1})^{\beta}_{\gamma}
{\chi}^{\mu}_B (\Phi ^{-1})^{\nu}_{\mu}
{\bar v}^{\alpha}_{\beta}{\bar v}^{\sigma}_{\nu}
\end{array}
\right).
\label{9}
\end{equation}
In (\ref{9}), ${\bar v}^{\sigma}_{\nu}\equiv 
{\bar v}^{\sigma}_{\nu}(a)$ and other components depend on $Q^{\ast}$.

The pseudoinversion of ${\tilde G}_{\cal B\cal C}$ means that
\begin{eqnarray*}
\displaystyle
{\tilde G}^{\cal A\cal B}{\tilde G}_{\cal B\cal C}=\left(
\begin{array}{cc}
(P_{\perp})^C_B & 0\\
0 & {\delta}^{\alpha}_{\beta}
\end{array}
\right).
\end{eqnarray*}

The determinant of the matrix (\ref{8}) is equal to
\begin{eqnarray*}
&&(\det {\tilde G}_{\cal A\cal B})=
\det G_{AB}(Q^{\ast})\det {\gamma}_{\alpha \beta}(Q^{\ast})
(\det {\chi}{\chi }^{\top})^{-1}(Q^{\ast})
(\det {\bar u}^{\mu}_{\nu}(a))^2\nonumber\\
&&\,\,\,\,\,\,\,\,\,\,\times(\det 
{\Phi}^{\alpha}_{\beta} (Q^{\ast}))^2
\det (P_{\perp})^C_B(Q^{\ast}).
\end{eqnarray*}
It does not vanish only on the surface $\{\chi =0\}$.
On this surface $\det (P_{\perp})^C_B$ is equal to  unity. 

\section{Transformation of the stochastic process\\ and the 
semigroup}
In result of the coordinate replacement, the local stochastic
processes $\eta ^A(t)$ on the principal fiber bundle,  will get their
new representations. Applying the methods of \cite{Dalecky} to
obtained local processes we can form a new global process $\zeta (t)$.
It means that we have performed the  transformation of the global
process $\eta (t)$ to the process $\zeta (t)$. 

The global process $\zeta (t)$ has two kind of the 
local components:\\
$(Q^{\ast}{}^A(t),a^{\alpha}(t))$. 
The components $a^{\alpha}(t)$ describe the part of the stochastic
evolution that originates from the stochastic evolution that was given
on the group $\cal G$.
The $Q^{\ast}{}^A(t)$---evolution has its origin in the stochastic
evolution given on the submanifold $\Sigma$.

Although the process $Q^{\ast}{}^A(t)$ is described in terms of  the
dependent coordinates, the transformation of  the local stochastic
process $\eta ^A(t)$ for the process $\zeta ^{\cal A}(t)=
(Q^{\ast}{}^A(t),a^{\alpha}(t))$ is, in fact, the phase space
transformation of the process $\eta ^A(t)$. It take place because the
variables $Q^{\ast}{}^A$ are constrained by the condition: $\chi
^{\alpha}(Q^{\ast})=0$ and it is  valid  for the stochastic processes
too. 

But it is known that the phase space transformation of the stochastic
processes does not change the probabilities. It means that the action
of the local semigroup
${\tilde U}_{\eta}$  on a function $\varphi _0 (p)$ is equal to the
expectation of the transformed function given
a $\sigma $--algebra generated by the transformed process 
$\zeta ^{\cal A}(t)$. 

On charts of the manifold $\cal P$, this transition to new coordinates
can be considered as follows.
The local semigroup 
\[
{\tilde U}_{\eta}(s,t) \phi _0 (p)=
{\rm E}_{s,p}\phi _0 (\eta (t)),\,\,\,
s\leq t,\,\,\,\eta (s)=p
\]
for the process $\eta (t)$ which is restricted to the chart 
$({\cal V}_p,\varphi ^{\cal P})$,
\[
\varphi ^{\cal P}(\eta (t))={\eta}^{\varphi 
^{\cal P}}(t)\equiv\{\eta ^A (t)\},
\]
 can be written as
 \[
 {\tilde U}_{\eta}(s,t) \phi _0 (p)={\rm E}_
 {s,\varphi ^{\cal P}(p)}
 \phi _0\left((\varphi ^{\cal P})^{-1}
 (\eta ^{\varphi ^{\cal P}}(t))\right),
 \,\,\,\eta ^{\varphi ^{\cal P}}(s)=
 \varphi ^{\cal P}(p).
 \]
The phase space transformation of the local stochastic processes
\[
\eta ^A(t)=F^A(Q^{\ast}{}^B(t),a^{\alpha}(t))
\]
transforms the local semigroup ${\tilde U}_{\eta}$: 
\[
{\tilde U}_{\eta}(s,t) \phi _0(p)={\rm E}_
{s,{\tilde{\varphi}}^{\cal P}(p)}
\phi _0\left(({\tilde{\varphi}}^{\cal P})^{-1}
({\zeta}^{{\tilde{\varphi}}^{\cal P}}(t))\right)=
{\rm E}_
{s,{\tilde{\varphi}}^{\cal P}(p)}
{\tilde{\phi _0}}
\left(
{\zeta}^{{\tilde{\varphi}}^{\cal P}}(t)\right),
\]
where $\left(
{\tilde{\varphi}}^{\cal P}
\right)^{-1}
=\left(\varphi ^{\cal P}\right)^{-1}
\circ F$ and ${\tilde{\phi _0}}={\phi _0}\circ
({\tilde{\varphi}}^{\cal P})^{-1}$.

Notice, that there is a local isomorphism of the principal fiber
bundle $P(\cal M,\cal G)$ and the trivial principal fiber bundle
$P_{\Sigma}=\Sigma\times {\cal G}
\rightarrow \Sigma$ \cite{Mitter,Kelnhofer}.
From this fact it follows that we can  introduce such  charts in the
atlas for the total space $\cal P$ of the principal fiber bundle
$P(\cal M,\cal G)$ that are  related with the submanifold $\Sigma$. 

Therefore,  in our local semigroups we should take the expectation
values with respect to the measures defined by  the probability
distribution of the local processes ${\zeta}^{{\tilde{\varphi}}^{\cal
P}}(t)=
(Q^{\ast}{}^A(t),a^{\alpha}(t))$.  If these processes are consistent
with each other on  overlapping of the charts, we can define, by the
method of \cite{Dalecky},  the global process and global semigroup 
In turns, the fact of the consistence of the local processes is
verified by studying the transformations of the local stochastic
differential equations that are used to define the local stochastic
processes. 

\section{Stochastic differential equations}
Let us consider the stochastic differential equation for the component
$Q^{*}{}^A$ of the local stochastic process    
 ${\zeta}^{\cal A}(t)=
(Q^{*}{}^A(t), a^{\alpha}(t))$.
We suppose that the stochastic differential equation for this variable
has the following form:
\begin{equation}
dQ^{*}{}^{A}(t)=b^{*}{}^{A}(t)dt
+c^{*}{}^{A}_{\bar{B}}(t)
dw^{\bar{B}}(t),
\label{sd1}
\end{equation}
where we should define explicitly the drift and the diffusion
coefficient. 

Being subjected to the constraint condition
$\chi ^{\alpha}(Q^{*})=0$, the coordinates  $Q^{*}{}^A$ are the
functions of $Q^A$:
\[
Q^{*}{}^A=F^A(Q,a^{-1}(Q)).
\]
The stochastic variable ${\eta}^A(t)$ will have the same dependence on
the stochastic variable $Q^{*}{}^A(t)$.  

Applying the It\^o differentiation formula to the stochastic variable
$Q^{*}{}^A(t)$ we rewrite the left--hand side of the equation
(\ref{sd1}) as follows:
\begin{equation}
dQ^{*}{}^A(t)=\frac{\partial Q^{*}{}^A}{\partial
Q^E}d{\eta}^E(t)
+\frac12\frac{{\partial}^2Q^{*}{}^A}{\partial Q^E 
\partial Q^C}<d{\eta}^E(t)d{\eta}^C(t)>.
\label{sd2}
\end{equation}

Then, putting an expression of the stochastic differential
$d{\eta}^A(t)$  from (\ref{3}) into the right--hand side of
(\ref{sd2}), we obtain:
\begin{eqnarray}
&&dQ^{*}{}^A(t)=\frac{\partial Q^{*}{}^A}
{\partial Q^E}
 \left(-\frac12{\mu}^2\kappa 
 G^{PB}({\eta}(t))\,\Gamma^E_{PB}({\eta}(t))dt\right.
 \nonumber\\
 &&\,\,\,+\left.{\mu}
 \sqrt{\kappa} 
 {\mathfrak X}^E_{\bar M}({\eta}(t))dw^{\bar M}(t)\right)
+\frac12\frac{{\partial}^2Q^{*}{}^A}{\partial Q^E 
\partial Q^C}<d{\eta}^E(t)d{\eta}^C(t)>,
\label{sd3}
\end{eqnarray}
 where ${\Gamma}^E_{PB}$ are the Christoffel coefficients for the
 Riemannian metric $G_{AB}$.
In order to express the stochastic variable ${\eta}^A(t)$ 
in the last equation
in terms of   $Q^{*}{}^A(t)$ and $a^{\alpha}(t)$ 
we make  use  the equation ${\eta}^A(t)=
F^A(Q^{*}{}^B(t),a^{\alpha}(t))$.

After such a transformation we find that the coefficient which stands
at the differential  $dt$ in the obtained expression will be the drift
of the equation (\ref{sd1}). And, correspondently, the term at the
stochastic differential $dw(t)$  will be  the diffusion coefficient.
As a result we obtain the following equation for $Q^{*}(t)$:
\begin{eqnarray}
&&dQ^{*}{}^A(t)=\frac12 {\mu}^2\kappa\Bigl[
N^A_CN^R_M
(G(Q^{*}{}(t))^{-1/2}\frac{\partial}
{\partial Q^{*}{}^R}\biggl(G(Q^{*}(t))^{1/2}G^{CM}
(Q^{*}(t))\biggr)\Bigr.
\nonumber\\
&&+N^A_{CL}G^{CL}-G^{PC}N^K_CK^M_{\mu
M}
(\Phi ^{-1})^{\mu}_{\nu}{\chi}^{\nu}_P+
G^{PC}N^A_CK^E_{\mu P}
(\Phi ^{-1})^{\mu}_{\nu}{\chi}^{\nu}_E
\nonumber\\
&&+\Bigl.	G^{PB}N^A_CK^C_{\mu B}
(\Phi ^{-1})^{\mu}_{\nu}{\chi}^{\nu}_P\Bigr]dt+
\mu\sqrt{\kappa}N^A_C \tilde{\mathfrak X}^C_{\bar M}(Q^{*}(t))
dw^{\bar M}(t).
\label{sd4}
\end{eqnarray}
In this equation all variables depend on $Q^{*}(t)$  
and by  additional lower indices we denote   the corresponding
derivatives. (For example, $N^A_{CL}(Q^{*})\equiv
\frac{\partial}{\partial Q^L}
N^A_C(Q)|_{Q=Q^{*}}$.)\\
Also, $\tilde\mathfrak X$ from (\ref{sd4}) is defined  by
$\sum^{n_{P}}_{\bar{
{\scriptscriptstyle K}}
\scriptscriptstyle =1}
\tilde{\mathfrak X}_{\bar{K}}^A(Q^{*})\tilde{\mathfrak X}_
{\bar{K}}^B(Q^{*})=G^{AB}(Q^{*})$.

The drift term of equation (\ref{sd4}) has another representation,
which is related with the geometrical objects that are specific for
the considered problem.
In order to the derive this representation we make use the following
expansion of the operator Laplace -- Beltrami from \cite{Kunstatter}:
\begin{equation}
 \frac12{\triangle}_{\cal
 P}=\frac12\Bigl({}^{\bot}G^{AB}
 {\nabla}_A{\nabla}_B-K^A_{\mu}{\gamma}^{\mu\nu}
 ({\nabla}_AK^B_{\nu}){\nabla}_B+K^A_{\alpha}
 {\gamma}^{\alpha\beta}{\nabla}_BK^B_{\beta}
 {\nabla}_B\Bigr),
 \label{sd5}
 \end{equation}
 where ${}^{\bot}G^{AB}=G^{AB}-K^A_{\alpha}{\gamma}
 ^{\alpha\beta}K^B_{\beta}$, and ${\nabla}_A$ 
 is the symbol of the covariant derivative
which is obtained from the Christoffel coefficients for the original
Riemannian metric $G_{AB}(Q)$.

If we replace $Q$ for $(Q^{\ast},a)$ in (\ref{sd5}), then it can be
shown that the drift in (\ref{sd4}) is  a sum of two terms: $b_{\Roman
1}^A(Q^{*})$ and $b_{\Roman 2}^A(Q^{*})$. 
They are equal to those coefficients at the first partial derivatives
over $Q^{*}$ that come from the first and the second terms of the
right--hand side of the equation (\ref{sd5}) after changing the
variables.
The third term  of expansion of the laplacian in (\ref{sd5}) does not
give the contribution to the terms with 
 partial derivatives over $Q^{*}$.

Performing  necessary evaluations, we find that  
$b_{\Roman 2}^A(Q^{*})$ is the projection 
of the mean curvature vector of the orbit
\[
j^D(Q)\frac{\partial}{\partial Q^D}=
\frac12{\Pi}^D_A(Q){\gamma}^{\alpha\beta}(Q)
\left[{\nabla}_{K_{\alpha}(Q)}
  K_{\beta}(Q)\right]^A\frac{\partial}{\partial Q^D}
\]
on the submanifold $\{\chi ^{\alpha}=0\}$.
The projection is given with the help of the transformed metric
${\tilde G}^{\cal A\cal B}$ 
 as follows:
\[
{\tilde G}^{SL}
  \tilde G\left(j^D\frac{\partial}{\partial Q^D},
  \frac{\partial}{\partial Q^{*}{}^S}\right)
  \frac{\partial}{\partial Q^{*}{}^L},
  \] 
where before taking the projection one should change the  variables
$Q$ in $j^D\frac{\partial}{\partial Q^D}$ for $Q^{*}$ and $a$.
The projection operator ${\Pi}^D_A={\delta}^D_A-K^D_{\mu}
{\gamma}^{\mu \nu}K^C_{\nu}G_{CA}$ extracts the direction which is
normal to the orbit: ${\Pi}^D_A K^A_{\alpha}=0$. 

As a result of the projection we get the following expression for the
$b_{\Roman 2}^A(Q^{*})$:
\[
  b_{\Roman
  2}^A(Q^{*})=\frac12G^{EU}N^A_EN^D_U
  \left[{\gamma}^{\alpha\beta}G_{CD}
  ({\tilde{\nabla}}_{K_{\alpha}}
  K_{\beta})^C\right],
  \]
in which all the values from the right--hand side  depend on $Q^{*}$
and by $({\tilde{\nabla}}_{K_{\alpha}}
  K_{\beta})^C(Q^{*})$ 
  we denote
\[
  K^A_{\alpha}(Q^{*})\left.\frac{\partial}
  {\partial Q^A}K^C_{\beta}(Q)\right |_{Q=Q^{*}}+
  K^A_{\alpha}(Q^{*})K^B_{\beta}(Q^{*})
  {\tilde \Gamma}^C_{AB}(Q^{*}),
    \]
  where 
\begin{eqnarray*}
&&
{\tilde \Gamma}^C_{AB}(Q^{*})=
\nonumber\\
&&\,\,\,\,\,\,\,
\frac12\
 G^{CE}(Q^{*})\left(\frac{\partial}
 {\partial {Q^{*}}^A}G_{EB}(Q^{*})+
 \frac{\partial}
 {\partial {Q^{*}}^B}G_{EA}(Q^{*})-
 \frac{\partial}
 {\partial {Q^{*}}^E}G_{AB}(Q^{*})\right).
 \end{eqnarray*}

The relation of $b_{\Roman 1}^A(Q^{*})$, which comes from the first
term of an expansion of the laplacian in  (\ref{sd5}), with the
geometry of the problem can be found  as follows.

In the local picture, the projection onto the orbit space $\cal M$,
which is locally isomorphic to $\Sigma$, is realized  by replacement
of the coordinates: $Q^A=Q^{\ast}{}^A(x)$.
Under this replacement
the first term in the right--hand side of equation   (\ref{sd5})
transforms into the Laplace -- Beltrami operator of the manifold
(${\cal M}, h_{ij}$) with the induced metric 
\[
h_{ij}(x)=Q^{*}{}^A_i(x)\,G^H_{AB}
 (Q^{*}(x))\,Q^{*}{}^B_j(x).
 \]
We can also regard the orbit space  as a submanifold of the
(Riemannian) manifold $({\cal P},G^H_{AB}(Q))$ with the degenerate
metric $G^H_{AB}$.

 The orbit space diffusion is   described locally by the following
 stochastic differential equation:
\[
dx^i(t)=-\frac12{\mu}^2\kappa h^{kl}(x(t))
{\Gamma}^i_{kl}(x(t))dt+\mu\sqrt{\kappa}
X^i_{\bar m}(x(t))dw^{\bar m}(t),
\]
in which the Christoffel coefficients correspond to the  metric
$h_{ij}(x)$.

But, besides the standard description of this diffusion in terms of
the internal variables that are given on the submanifold, there is a
description of the same diffusion with the help of the stochastic
differential equation  defined in terms of the variables of the
external manifold. In \cite{Lewis}, it was considered the particular
case of such a description when it was used the Euclidean space  as an
external manifold. It is not difficult to find a similar discription
for a general case (see Appendix A). 

To derive the corresponding stochastic differential equation of our
problem one should repeate the evaluation that was done in  Appendix
A. 

We remark, that in our case the metric $G^H_{AB}$ of an external
manifold is degenerated one. Therefore, instead of the relation
(\ref{ap4}) from Appendix A we will have  
\[
 h^{kl}(x){\Gamma}^i_{kl}=G^{H}_{AB}
 \left(
 Q^{*}{}^A_{kl}h^{
 kl}
 +{}^H{\Gamma}^A_{CD}
 Q^{*}{}^C_{k}Q^{*}{}^D_l
 h^{kl}
 \right)
 h^{im}Q^{*}{}^B_{m},
 \]
 where the multiplication
 $G^{H}_{AB}(Q^{*}(x))\,\,{}^H{\Gamma}^B_{CD}(Q^{*}(x))$ 
 is defined as
\begin{equation}
G^{H}_{AB}\,\,
{}^H{\Gamma}^B_{CD}
=\frac12\left(G^{H}_{AC,D}+
G^{H}_{AD,C}-G^{H}_{CD,A}
\right).
\label{sd6}
\end{equation}
In (\ref{sd6}), by the derivatives we mean the following: 
$G^{H}_{AC,D}\equiv 
\left.{{\partial G^{H}_{AC}(Q)}\over
{\partial Q^D}}\right|_{Q=Q^{*}(x)}$.

 Repeating all the steps that was done in the  Application we can show
 that in result of the replacement of the variables
the  drift of the obtained stochastic differential equation coincides
with  the coefficient which stands 
at the first partial derivative over $Q^*$
 in the term that arise  from the first term of the expansion of the
 laplacian in (\ref{sd5}). 
 Hence, this drift coincides also with $b_{\Roman 1}$. The last means
 that $b_{\Roman 1}$ should  be related with the geometric values that
 characterize the orbit space.

Performing the aforementioned  transformations we get  the following
expression for $b_{\Roman 1}$:
\[
b^{A}_{\Roman
1}\left(Q^{*}(x)\right)=-\frac12G^{EM}\left
(Q^{*}(x)\right)
N^C_E\left(Q^{*}(x)\right)
N^B_M\left(Q^{*}(x)\right)
{}^H{\Gamma}^A_{CB}\left(Q^{*}(x)\right)+
j^{A}_{\Roman 1}\,,
\]
where $j_{\Roman 1}$  is the mean curvature vector of the orbit space.
It can be evaluated as follows:
\[
j^A_{\Roman 1}=\frac12\left({\delta}^A_B-
N^A_B\left(Q^{*}(x)\right)\right)
h^{ij}(x)
\left[Q^{*}{}^B_{ij}+{Q^*}{}^P_i{Q^*}{}^L_j\,
{}^H{\Gamma}^B_{PL}\left(Q^{*}(x)\right)
\right].
\]

 But, as a function, the mean curvature    is given on a submanifold.
 So, similarly to that as was done in the Application A,  we can
 redefine  the stochastic variable $Q^{*}(x(t))$ 
 for a new stochastic variable $Q^{*}(t)$. (We denote a new stochastic
 variable  by the same latter.) 
  
  Notice, that from equation (\ref{sd6}) the Christoffel symbols
  ${}^H{\Gamma}^B_{CD}$ are defined up to the terms $T^B_{CD}$ that
  are satisfied to $G^H_{AB}T^B_{CD}=0$. However, this ambiguity is
  not essential, since $b_{\Roman 1}$ can be also presented as
  \[
  b^{A}_{\Roman 1}=-\frac12 N^A_E
  \,{}^H{\Gamma}^E_{CD}
  N^C_K N^D_UG^{KU}+\frac12 N^A_{LM}N^L_K
  N^M_UG^{KU}.
  \]

  Therefore, in result of the transformation of equation  (\ref{sd1})
  we get the following stochastic differential equation:
  \begin{equation}
dQ^{*}{}^{A}(t)={\mu}^2\kappa
\biggl(-\frac12
G^{EM}N^C_EN^B_M\,{}^H{\Gamma}^A_{CB}+
j^{A}_{\Roman 1}+j^{A}_{\Roman
2}\biggr)dt +\mu\sqrt{\kappa}
N^A_C\tilde{\mathfrak X}^C_{\bar M}dw^{\bar M},
\label{sd8}
\end{equation}
where all the values from the right--hand side now depend on
$Q^{*}(t)$ and we have introduced a new notation for $b^A_{\Roman 2}$.
In new notation it is denoted by $j^A_{\Roman 2}$.
  
  As for the stochastic differential equation for the group variable
  $a^{\alpha}(t)$, it can be obtained by the same method as it was
  done for the  variable $Q^*{}^A(t)$.
    It can be found that this equation is the following:
  \begin{eqnarray}
  &&da^{\alpha}=-\frac12{\mu}^2\kappa\biggl[G^{RS}
  \tilde{\Gamma}^B_{RS}(Q^*){\Lambda}^{\beta}_B
  {\bar v}^{\alpha}_{\beta}
  +G^{RP}{\Lambda}^{\sigma}_R
  {\Lambda}^{\beta}_BK^B_{\sigma P}
  {\bar v}^{\alpha}_{\beta}
  \nonumber\\
  &&\,\,\,\,\,\,\,\,\,\,-G^{CA}N^M_C\frac{\partial}
  {\partial Q^*{}^M}
  \biggl({\Lambda}^{\beta}_A\biggr)
  {\bar v}^{\alpha}_{\beta}
  -G^{MB}{\Lambda}^{\epsilon}_M
  {\Lambda}^{\beta}_B{\bar v}^{\nu}_{\epsilon}
  \frac{\partial}
  {\partial a^{\nu}}
  \bigl({\bar v}^{\alpha}_{\beta}\bigr) 
  \biggr]dt\nonumber\\
  &&\,\,\,\,\,\,\,\,\,+\mu\sqrt{\kappa}
  {\bar v}^{\alpha}_{\beta}{\Lambda}^{\beta}_B
  \tilde{\mathfrak X}^B_{\bar M}dw^{\bar M},
  \label{sd10}
  \end{eqnarray}
where ${\bar v}\equiv {\bar v}(a)$ and 
other coefficients depend on $Q^*$.
Also, we have introduced a new notation:  
\[
{\Lambda}^{\alpha}_B=({\Phi}^{-1})^{\alpha}_{\mu}
{\chi}^{\mu}_B.
\]
  In (\ref{sd10}), the Christoffel symbols
  $\tilde{\Gamma}^B_{RS}(Q^*)$ are obtained from ${\Gamma}^A_{BC}(Q)$,
  if in its definition we rewrite the derivatives by the formula
  (\ref{7}).
  
  Therefore, the stochastic process $\zeta (t)$ is given locally by
  the solution of the stochastic differential equations (\ref{sd8})
  and (\ref{sd10}).
  On charts of the manifold the set of the solutions of these
  equations determine the local stochastic evolution families of
  mappings of the manifold $\cal P$.
  
  As in \cite{Dalecky}, with these local families  it is possible to
  define the global stochastic process $\zeta (t)$ which consist of
  two components  related with the stochastic evolution on the
  submanifold $\Sigma $ (the gauge surface) and with the stochastic
  evolution on the orbit of the principal fiber bundle.
  
  The performed transformation of the stochastic process $\eta(t)$
  results to the corresponding transformation of the global semigroup
  (\ref{5}). 
    Now our semigroup is determined by the superposition of the local
    semigroups ${\tilde U}_{{\zeta}^{\varphi ^P}}$:
  \begin{equation}
\psi _{t_b}(p_a,t_a)=
{\lim}_q {\tilde U}_{{\zeta}^{\varphi
^P}}(t_a,t_1)\cdot\ldots\cdot
{\tilde U}_{{\zeta}^{\varphi ^P}}
(t_{n-1},t_b) 
{\tilde \phi} _0(Q^*_a, \theta _a),
\label{sd21}
\end{equation}
where 
\[
{\tilde U}_{{\zeta}^{\varphi ^P}}(s,t) 
{\tilde \phi}_0 (Q^*_0,\theta_0)={\rm E}_
{s,(Q^*_0,\theta _0)}
\tilde{\phi}_0(Q^*(t),a(t)),
\,\,\,Q^*(s)=Q^*_0,\,\,\,a(s)=\theta _0.
\]
  We will write this  global semigroup in the following symbolical
  form: 
  \[
  {\psi}_{t_b} (p_a,t_a)={\rm E}\Bigl[\tilde{\phi
  }_0({\xi}_{\Sigma}(t_b),a(t_b))\exp
  \{\frac 1{\mu ^2\kappa m}\int_{t_a}^{t_b}
  \tilde{V}({\xi}_{\Sigma}(u))du\}\Bigr],
  \]
where ${\xi}_{\Sigma} (t_a)=Q^{*}_a$,
$a(t_a)=\theta _a$ , 
$\varphi ^P(p_a)=(Q^{*}_a,\theta _a)$ and
we have taken into account the omitted potential term.
  
  From (\ref{sd8}) and (\ref{sd10}) it follows that the coordinate
  representation of the differential generator of the semigroup
  related with the stochastic process $\zeta (t)$ is given by 
  \begin{eqnarray*}
&&\frac12{\mu}^2\kappa\left(G^{CD}N^A_CN^B_D
\frac{{\partial}^2}{\partial Q^{*}{}^A
\partial Q^{*}{}^B}-G^{CD}N^A_CN^B_D\,
{}^H{\Gamma}^E_{AB}
\frac{\partial}{\partial Q^{*}{}^E}
+j^A_{\Roman 1}
\frac{\partial}{\partial Q^{*}{}^A}\right.
\nonumber\\
&&+j^A_{\Roman 2}
\frac{\partial}{\partial Q^{*}{}^A}
+G^{AB}
{\Lambda}^{\alpha}_A{\Lambda}^{\beta}_B
{\bar L}_{\alpha}{\bar L}_{\beta}
-G^{RS}{\tilde {\Gamma}^B_{RS}}
{\Lambda}^{\alpha}_B{\bar L}_{\alpha}
-G^{RP}
{\Lambda}^{\sigma}_R {\Lambda}^{\alpha}_B
K^B_{\sigma P} {\bar L}_{\alpha}\nonumber\\
&&+G^{CA}N^M_C
\frac{\partial}{\partial Q^{*}{}^M}
\left({\Lambda}^{\alpha}_A\right)
{\bar L}_{\alpha}
\left. +2G^{BC}N^A_C{\Lambda}^{\alpha}_B
{\bar L}_{\alpha}\frac{\partial}
{\partial Q^{*}{}^A}\right)+\frac{1}
{{\mu}^2\kappa m}\tilde V.
\nonumber\\
\end{eqnarray*}
Here,  all the values, except  for ${\bar L}$, depend on $Q^{*}$.

  \section{Factorization of the path integral measure}
  In  \cite{Stor1,Stor2}, a new method of factorization of the path
  integral measure was proposed.  
  Here, we will apply it to our case of constrained (by the equation
  ${\chi}^{\alpha} (Q^{*})=0$) integration variables.
  
The main idea of \cite{Stor1} was in exploiting the stochastic
differential equation from the nonlinear filtering  theory
\cite{Lipster,Pugachev}. This equation describes the evolution of the
conditional mathematical expectation of the signal process ( the
process $a(t)$ in our case) with respect to the $\sigma$--algebra
generated by an observable process (the stochastic process $Q^*(t)$).

 In order to make use of this equation  we transform each local
 semigroup ${\tilde U}_{{\zeta}^{\varphi ^P}}$ from   (\ref{sd21}) as
 follows:
 \begin{equation}
{\tilde U}_{{\zeta}^{\varphi ^P}}(s,t) {\tilde
\phi} (Q^{*}_0,\theta _0)=
{\rm E}
\Bigl[{\rm E}\bigl[\tilde{\phi }(Q^{*}(t),a(t))\mid
(
{\cal F}_{Q^{*}})_{s}^{t}\bigr]\Bigr].
\label{25}
\end{equation}
The above transformation is based on the properties of the conditional
expectation of the Markov processes.
  Such a path integral transformation can be also regarded as an
  analog of the transition from the multiple integrals to the repeated
  ones in the ordinary integration. 
 
 Being the integrand  of  the "repeated" integral, the conditional
 expectation 
 \[
 \hat{\widetilde{\phi }}(Q^{*}(t))\equiv 
 {\rm E}\Bigl[\tilde{\phi }(Q^{*}(t),a(t))\mid (
 {\cal F}_{Q^{*}})_{s}^{t}\Bigr],
\]
 should satisfy the nonlinear filtering equation. With account of our
 stochastic differential equations (\ref{sd8}) and (\ref{sd10}), we
 get that it will be as follows:
 \begin{eqnarray}
&&d\hat{\widetilde{\phi }}=-\frac12{\mu}^2\kappa
\Bigl(
G^{RS}
\tilde{\Gamma}^B_{RS}
{\Lambda}^{\beta}_B+
G^{RP}{\Lambda}^{\sigma}_R{\Lambda}^{\beta}_B
K^B_{P\sigma}
-
G^{CA}N^M_C
\frac{\partial}{\partial Q^{*}{}^M}
({\Lambda}^{\beta}_A)\Bigr)
\nonumber\\
&&\times{\rm E}
[\bar{L}_\beta \tilde{\phi }
\mid ({\cal F}_{Q^{*}})_{s}^t]dt
+\frac12{\mu}^2\kappa
G^{CB}{\Lambda}^{\nu}_C{\Lambda}^{\kappa}_B
{\rm E}
[\bar{L}_\nu\bar{L}_{\kappa} \tilde{\phi }
\mid ({\cal F}_{Q^{*}})_{s}^t]dt
\nonumber\\
&&+{\mu}\sqrt{\kappa}{\Lambda}^{\beta}_C{\Pi}^C_K\tilde{\mathfrak
X}^K_{\bar
M}
{\rm E}
[\bar{L}_\beta \tilde{\phi }
\mid ({\cal F}_{Q^{*}})_{s}^t]dw^{\bar M},
\label{26}
\end{eqnarray} 

 Further transformation of the equation (\ref{26})
 consists in separation of the space variables from the group ones.
 It can be done by applying the Peter -- Weyl theorem 
 to the function ${\widetilde{\phi }}$ considered as the function on a
 group $\cal G$. 
 An expansion of this function in a series  yields 
\[
\tilde{\phi }(Q^{*},a)=\sum_{\lambda
,p,q}c_{pq}^\lambda
(Q^{*})D_{pq}^\lambda (a)\,,
\]
where $D^{\lambda}_{pq}(a)$ are 
    the matrix elements  of an irreducible representation
    $T^{\lambda}$ of the group $\cal G$:
 $\sum_qD_{pq}^\lambda
(a)D_{qn}^\lambda (b)=D_{pn}^\lambda (ab)$.
 
 By   the properties of  the conditional mathematical expectations we
 have  
 \begin{eqnarray*}
 {\rm E}\bigl[\tilde{\phi }(Q^{*}(t),a(t))\mid
 ({\cal F})
 _{Q^{*}})_{s}^t\bigr]&=&\sum_{\lambda
 ,p,q}c_{pq}^\lambda (Q^{*}(t))\,
 {\rm E}\bigl[D_{pq}^\lambda
 (a(t))\mid ({\cal F}_{Q^{*}})_{s}^t\bigr]
 \nonumber\\
 &\equiv&
 \sum_{\lambda ,p,q}c_{pq}^\lambda (Q^{*}(t))\,
 \hat{D}_{pq}^\lambda(Q^{*}(t)),
 \end{eqnarray*}
where 
\[
c_{pq}^\lambda (Q^{*}(t))=d^\lambda \int_{\mathcal G}\tilde{\varphi }
(Q^{*}(t),\theta ) 
{\bar D}_{pq}^\lambda (\theta )d\mu (\theta ),
\]
($d^{\lambda}$ is a dimension of an irreducible representation
and 
$d\mu (\theta )$ is a normalized ($\int_{\mathcal G}d\mu (\theta
)=1$)
invariant Haar measure on a group $\mathcal G$).
 
 Then, in a similar manner as it was done in \cite{Stor1,Stor2}
 one can derive the stochastic differential equation for the
 conditional expectation $\hat{D}_{pq}^\lambda$:
 \begin{eqnarray}
&&d\hat{D}_{pq}^\lambda (Q^{*}(t))=
\nonumber\\
&&-\frac12{\mu}^2\kappa\left\{\Bigl[
G^{RS}
\tilde{\Gamma}^B_{RS}
{\Lambda}^{\mu}_B+
G^{RP}{\Lambda}^{\sigma}_R{\Lambda}^{\mu}_B
K^B_{P\sigma}
-\Bigl. G^{CA}N^M_C
\frac{\partial}{\partial Q^{*}{}^M}
({\Lambda}^{\beta}_A)\Bigr]\right.
\nonumber\\
&&\left.\times (J_\beta)_{pq^{\prime
}}^\lambda \hat{D}_{q^{\prime }q}^\lambda
(Q^{*}(t))
-
G^{CB}{\Lambda}^{\alpha}_C
{\Lambda}^{\nu}_B
\,\,(J_\alpha)_{pq^{\prime }}^\lambda (J_\nu
)_{q^{\prime }q^{\prime \prime }}^\lambda
\hat{D}_{q^{\prime \prime
}q}^\lambda (Q^{*}(t))\right\}dt
\nonumber\\
&&+\mu\sqrt{\kappa}{\Lambda}^{\nu}_C{\Pi}^C_K
(J_\nu )_{pq^{\prime }}^\lambda \hat{D}_{q^{\prime
}q}^\lambda
(Q^{*}(t))\tilde{\mathfrak X}_{\bar{M}}^K(Q^{*}(t))
dw^{\bar{M}}(t),
\label{28}
\end{eqnarray}
 in which $\left.(J_\mu )_{pq}^\lambda \equiv (\frac{\partial
D_{pq}^\lambda (a)}
{\partial a^\mu })\right|_{a=e}$ are the infinitesimal generators of
the representation $D^{\lambda}(a)$:
\[
\bar{L}_\mu D_{pq}^\lambda (a)=\sum_{q^{\prime
}}(J_\mu )_{pq^{\prime
}}^\lambda D_{q^{\prime }q}^\lambda (a).
\] 

  Notice, that in general, the conditional expectation
  $\hat{D}_{pq}^\lambda (Q^{*}(t))$ depend also on the initial points
  $Q^{*}_0=Q^{*}(s)$ and $\theta^{\alpha}_0=a^{\alpha}(s)$ besides the
  of the stochastic processes $Q^{*}(t)$. 
 
 The solution of the linear matrix  stochastic differential equation
 (\ref{28}) can be written \cite{Dalmulti} as:
 \begin{equation}
\hat{D}_{pq}^\lambda(Q^{*}(t))=
(\overleftarrow{\exp })_{pn}^\lambda
(Q^{*}(t),t,s)\,{\rm E}\bigl[D_{nq}^\lambda
(a(s))\mid ({\cal F}_{Q^{*}})_{s}^t\bigr],
\label{29}
\end{equation}
where 
 \begin{eqnarray}
&&(\overleftarrow{\exp })_{pn}^\lambda
(Q^{*}(t),t,s)=\overleftarrow{\exp
}%
\int_{s}^t\Bigl\{\frac 12{\mu}^2\kappa
\Bigl[\bar{\gamma }^{\sigma
\nu }(Q^{*}(u))(J_\sigma
)_{pr}^\lambda (J_\nu )_{rn}^\lambda 
\nonumber\\
&&-\biggl(
G^{RS}
\tilde{\Gamma}^B_{RS}
{\Lambda}^{\beta}_B
+
G^{RP}{\Lambda}^{\sigma}_R{\Lambda}^{\beta}_B
K^B_{P\sigma}
-G^{CA}N^M_C
\frac{\partial}{\partial Q^{*}{}^M}
({\Lambda}^{\beta}_A)\biggr)
\,\,(J_\beta)_{pn}^\lambda \Bigr]du
\nonumber\\
&&+\mu\sqrt{\kappa}{\Lambda}^{\beta}_C(J_\beta)_{pn}^\lambda 
{\Pi}^C_K\tilde{\mathfrak X}^K_{\bar M}dw^{\bar M}
\Bigr\}
\label{30}
\end{eqnarray}
is a multiplicative stochastic integral. This integral is a limit of
the sequence of time--ordered multipliers that have been obtained as a
result of breaking of a time interval $[s,t]$. The time order of these
multipliers is given by the direction of the arrow aimed to the
multipliers at greater times.
 
 With account of the representation for $\hat{D}_{pq}^\lambda $
 obtained in (\ref{29}) and (\ref{30}) we rewrite our local semigroup
 (\ref{25}) as follows:
 \begin{equation}
{\tilde U}_{{\zeta}^{\varphi ^P}}(s,t) {\tilde
\phi} (Q^{*}_0,\theta _0)
=\sum_{\lambda ,p,q,q^{\prime }}{\rm E}
\bigl[
c_{pq}^\lambda (Q^{*}(t))
(\overleftarrow{\exp })_{pq^{\prime }}^\lambda 
(Q^{*}(t),t,s)\bigr] D_{q^{\prime}q}^\lambda
(\theta
_0),
\label{31}
\end{equation}
where we have taken into account that 
 \[
{\rm E}\bigl[D_{nq}^\lambda (a(s))\mid ({\cal
F}_{Q^{*}})_{s}^t\bigr]
=D_{nq}^\lambda (a(s))=D_{nq}^\lambda (\theta _0).
\]
 
 In order to obtain the global semigroup by the methods of
 \cite{Dalecky} one should break the time interval $[t_a,t_b]$ and
 should take the superposition of the local semigroups that are
 similar to (\ref{31}). 
 Then, the global semigroup for the global process is obtained as a
 limit (under the refinement of the subdivision of the time interval)
 of 
 the superposition of these local semigroups. 
 The relation between the global semigroup obtained in result of the
 limiting procedure can be  written simbolically in the following
 form:
 \begin{equation}
 {\psi}_{t_b}(p_a,t_a)
 =\sum_{\lambda ,p,q,q^{\prime }}{\rm E}
 \bigl[
  c_{pq}^\lambda ({\xi}_{\Sigma}(t_b))
 (\overleftarrow{\exp })_{pq^{\prime }}^\lambda 
 ({\xi}_{\Sigma}(t),t_b,t_a)\bigr]
 D_{q^{\prime}q}^\lambda (\theta _{a})
 \label{32}
 \end{equation}
$({\xi}_{\Sigma}(t_a)={\pi}|_{\Sigma}\circ p_a$),   \\ 
where ${\xi}_{\Sigma}(t)$ is a global stochastic process defined on
the submanifold $\Sigma$. This process is described locally by the
equations (\ref{sd8}).
 
 Thus, our initial original path integral has been rewritten as the
 sum of the matrix semigroups (the path integrals) that are given on
 the submanifold $\Sigma$.
 The differential generator (the Hamiltonian operator) of  these
 matrix semigroups is 
 \begin{eqnarray}
 &&\frac12\mu ^2\kappa \left\{\left[
 G^{CD}N^A_CN^B_D\frac{{\partial}^2}{\partial
 Q^{*}{}^A\partial
 Q^{*}{}^B}-G^{CD}N^E_CN^M_D\,{}^H{\Gamma}^A_
 {EM}\frac{\partial}{\partial
 Q^*{}^A}
 \right.\right.
 \nonumber\\
 &&+\left.\left(j^A_{\Roman1}+j^A_{\Roman
 2}\right)\frac{\partial}{\partial Q^*{}^A}
 \right](I^\lambda )_{pq}
 +2N^A_CG^{CP}{\Lambda}^{\alpha}_P
 (J_\alpha )_{pq}^\lambda
 \frac{\partial}{\partial Q^{*}{}^A}
 \nonumber\\
 &&-\left(G^{RS}{\tilde
 {\Gamma}}^B_{RS}{\Lambda}^{\alpha}_B+G^{RP}{\Lambda
 }^{\sigma}_R{\Lambda}^{\alpha}_BK^{B}_{P\sigma}-G^{
 CA}N^M_C\frac{\partial}{\partial
 Q^*{}^M}({\Lambda}^{\alpha}_A)\right)
 (J_\alpha )_{pq}^\lambda 
 \nonumber\\
 &&+\biggl. G^{SB}{\Lambda}^{\alpha}_B
 {\Lambda}^{\sigma}_S
 (J_\alpha)_{pq^{\prime }}^\lambda 
 (J_\sigma)_{q^{\prime }q}^\lambda \biggr\}
 \label{op2}
 \end{eqnarray} 
where $(I^\lambda )_{pq}$ is a unity matrix.
 
 The operator acts in the space of the section ${\Gamma}(\Sigma,V^*)$
 of the covector fiber bundle, which is associated with the trivial
 principal fiber bundle $\pi: \Sigma \times {\cal G}
\rightarrow \Sigma$. The scalar product in this space of the sections
is defined as follows:
 \begin{equation}
(\psi _n,\psi _m)=\int_{\Sigma}\langle \psi _n,\psi _m{\rangle}_
{V^{\ast}_{\lambda}}
 \frac{\det{\Phi}^{\alpha}_{\beta} }
 {{\det}^{1/2}({\chi}^{\mu}_A
 G^{AB}{\chi}^{\nu}_B)}\,dv_{\Sigma},
\label{33}
\end{equation}
 where $dv_{\Sigma}$  is the Riemannian volume element of $\Sigma $.

 An integration measure of the scalar product of the formula
 (\ref{33}) has been obtained from the Riemanian volume element of the
 manifold  $\cal P$, in which  we have changed the variables $Q^A$ for
 $(Q^*{}^i,
a^{\alpha})$. Also, we have used an equality:
 \begin{eqnarray*}
 &&
 \det G_{AB}(Q^*{}^i,
 Q^*{}^{\alpha}(Q^*{}^i),a^{\mu})=
 \nonumber
 \\
 &&\,\,\,\,\,\,\,\,\,\,\,\,\,\,\,\,
 \det \left((G_{\Sigma})_{AB}\right)
 \,\,{\det}^{-1}
 \left(
 (G^{BC}{\chi}^{\nu}_B{\chi}^{\mu}_C)
 {\Phi}^{-1}{}^{\alpha}_{\mu}
 {\Phi}^{-1}{}^{\beta}_{\nu} 
 {\bar v}^{\sigma}_{\alpha}
 {\bar v}^{\rho}_{\beta}
 \right).
 \end{eqnarray*}
 Notice, that the metric $(G_{\Sigma})_{AB}(Q^*{}^i,
Q^*{}^{\alpha}(Q^*{}^i))$
is a restriction of the metric $(P_{\bot})^C_A(Q^{*}) 
G_{CD}(Q^{*})(P_{\bot})^D_B(Q^{*})$ 
 to  the surface ${\Sigma}$.
 
 Performing the transformation of the measure in the integral
 (\ref{33}), we can also to present the scalar product as 
 \[
\int \langle \psi _n,\psi _m
{\rangle}_{V^{\ast}_{\lambda}}
\det{\Phi}^{\alpha}_{\beta}
\prod_{\alpha =1}^{N_{\cal G}}
\delta({\chi}^{\alpha}(Q^{*})){\det}^{1/2}G_{AB}
 \,dQ^{*}{}^1\wedge\dots\wedge 
dQ^{*}{}^{N_{\cal P}}.
 \]

 By taking an inversion  of (\ref{32}) we get the representation of
 standing under the sign of sum semigroups (the path integrals)
  in terms of the semigroups that are given on the manifold $\cal P$. 
 As in \cite{Stor2}, we will do it for the kernels of the
 corresponding local semigroups. 
  
Provided that the analytical restrictions are fulfilled,  the
semigroup from the left--hand side of (\ref{32}) ) can be presented as
 \begin{equation}
 {\psi}_{t_b} (p_a,t_a)=\int G_{\cal
 P}(p_b,t_b;p_a,t_a)\phi_0(p_b)dv_{\cal P}(p_b).
 \label{fac5}
\end{equation}

  Using the partition of the unity subordinated to a local finite
  covering of the manifold $\cal P$ and having in mind that there is a
  local isomorphism of $P_{\Sigma}(\Sigma, \cal G)$ with the trivial
  principal fiber  bundle $\varphi ^{\Sigma}_{\alpha _b}
({\cal U}^{\Sigma}_{\alpha _b})\times{\cal G}$,  
 by which a chart of the atlas of the manifold $\cal P$ is changed for
 the chart $\varphi ^{\Sigma}_{\alpha _b}
({\cal U}^{\Sigma}_{\alpha _b})\times{\cal G}$, we get the following
expression for the right--hand side of (\ref{fac5}):
  \begin{equation}
  \int\limits _{
  \varphi ^{\Sigma}_{\alpha _b}
  ({\cal U}^{\Sigma}_{\alpha _b})\times{\cal G}
  }
  \!\!\!\!\!\!\!\!\!\!\tilde{\tilde {\mu}}_{\alpha _b}(x_b)
   G_{\cal P}(\alpha _b,F(Q^{*}_b,\theta _b),
   t_b;
  \beta _a,F(Q^*_a,\theta _a),t_a)
  \tilde{\phi _0}(Q^*_b,\theta _b)dv(Q^*_b)
  d\mu(\theta _b),
  \label{34}
  \end{equation}
where $dv(Q^*)$ is the same volume measure as in (\ref{33}) 
and\\ $ d
\mu(\theta)= \det \bar{u}^{\alpha}_{\beta}
(\theta)\,d{\theta}^1\dots d{\theta}^
{N_{\cal G}}$ is a Haar measure on a group $\cal G$.
 
 Also, the right--hand side of (\ref{32}) can be presented locally as
\begin{equation}
\sum_{\alpha _b}\int\limits _{
\varphi ^{\Sigma}_{\alpha _b}
({\cal U}^{\Sigma}_{\alpha _b})
}
\tilde {\rho}_{\alpha _b}(x_b)
\sum_{\lambda ,p,q,q^{\prime }}
G^{\lambda}_{q^{\prime} p}(\alpha _b,Q^*_b,t_b;
\beta _a,Q^*_a,t_a)
 c_{pq}^\lambda (Q^*_b)
D_{q^{\prime}q}^\lambda (\theta _{a}) 
dv(Q^*_b).
\label{35}
\end{equation}

  Comparing (\ref{34}) and (\ref{35}) we find the relation between the
  local Green functions:
\begin{eqnarray*}
&&\displaystyle\int _{\cal G}
G_{\cal P}(\alpha _b,F(Q^*_b,
\theta _b),t_b;
\beta _a,F(Q^*_a,\theta _a),t_a) 
D_{pq}^\lambda (\theta _{b})d\mu (\theta _b)=\nonumber\\
&&
 \sum_{q^{\prime }}
G^{\lambda}_{q^{\prime} p}(\alpha _b,Q^*_b,t_b;
\beta _a,Q^*_a,t_a)
 D_{q^{\prime}q}^\lambda (\theta _{a}),
\end{eqnarray*}
which, with account of the unimodularity of the group $\cal G$, can be
easily reversed:
\begin{equation}
G^{\lambda}_{mn}(\alpha _b,Q^*_b,t_b;
\beta _a,Q^*_a,t_a)=
\displaystyle\int _{\cal G}G_P(\alpha _b,Q^*_b,\theta ,t_b;
\beta _a,Q^*_a,e,t_a) 
D_{nm}^\lambda (\theta )d\mu (\theta ).
\label{36}
\end{equation}
In this formula $e$ corresponds to the unity element of the group
$\cal G$ and
\[
 G_P(\alpha _b,Q^*_b,\theta _b,t_b;
 \beta _a,Q^*_a,\theta _a,t_a)\equiv
 G_{\cal P}(\alpha _b,F(Q^*_b,\theta _b),t_b;
\beta _a,F(Q^*_a,\theta _a),t_a).
\]

Since in the paper we have confined ourselves by the case of the
trivial principal fiber bundle, then the gluing these local Green
functions to the global Green functions can be done with the
transition coordinate functions defined for the charts of the
manifolds.

Hence, the equality (\ref{36}) can be extended from the local charts
to the whole manifold  and in result we obtain the relation  between
the Green function defined on the global  manifolds:
\begin{equation}
G^{\lambda}_{mn}(\pi _{\Sigma} (p_b),t_b;
\pi (p_a),t_a)=
\displaystyle\int _{\cal G}G_{\cal P}(p_b\theta,t_b;
p_a,t_a) 
D_{nm}^\lambda (\theta )d\mu (\theta ).
\label{38}
\end{equation}

The path integral from the left--hand side of this equality can be
written symbolically as 
\begin{eqnarray}
&&G^{\lambda}_{mn}(\pi _{\Sigma}(p_b),t_b;
\pi _{\Sigma}(p_a),t_a)=
\nonumber\\
&&{\tilde {\rm E}}_{{\xi _{\Sigma} (t_a)=
\pi _{\Sigma}(p_a)}\atop  
{\xi _{\Sigma}(t_b)=\pi _{\Sigma}(p_b)}}
\left[(\overleftarrow{\exp })_{mn}^\lambda 
(\xi _{\Sigma}(t),t_b,t_a)
\exp \left\{\frac 1{\mu ^2\kappa m}
\int_{t_a}^{t_b}\tilde{V}(\xi _{\Sigma}(u))
du\right\}\right]
\nonumber\\
&&=\int\limits_{{\xi _{\Sigma}(t_a)=
\pi _{\Sigma}(p_a)}\atop  
{\xi _{\Sigma}(t_b)=\pi _{\Sigma}(p_b)}} 
d{\mu}^{{\xi}_{\Sigma}}
\exp \left\{\frac 1{\mu ^2\kappa
m}\int_{t_a}^{t_b}\tilde{V}(\xi _{\Sigma}(u))
du\right\}
\nonumber\\
&&\times\overleftarrow{\exp}
\int_{t_a}^{t_b}\Bigl\{\frac 12{\mu}^2\kappa
\Bigl[{\gamma }^{\sigma
\nu }(\xi _{\Sigma}(u))(J_\sigma
)_{mr}^\lambda (J_\nu )_{rn}^\lambda 
\nonumber\\
&&-\biggl(
G^{RS}
\tilde{\Gamma}^B_{RS}
{\Lambda}^{\beta}_B+
G^{RP}{\Lambda}^{\sigma}_R{\Lambda}^{\beta}_B
K^B_{P\sigma}-G^{CA}N^M_C
\frac{\partial}{\partial Q^{*}{}^M}
({\Lambda}^{\beta}_A)\biggr)
\,\,(J_\beta)_{mn}^\lambda \Bigr]du
\nonumber\\
&&+\mu\sqrt{\kappa}{\Lambda}^{\beta}_C(J_\beta)_{mn}^\lambda 
{\Pi}^C_K\tilde{\mathfrak X}^K_{\bar M}dw^{\bar M}
\Bigr\}.
\label{39}
\end{eqnarray}

The semigroup defined by this kernel acts in the space of the
equivariant functions: 
\[
{\tilde \psi}_n (pg) =
D_{mn}^\lambda (g)
{\tilde \psi}_m (p),
\]
that are isomorphic to the functions $\psi _n$  from the space of the
sections $\Gamma (\Sigma, V^*)$ of the associated covector bundle:
\[
{\tilde \psi}_n (F(Q^{*},e))={\psi}_n (Q^*).
\]

 The method by which we have obtained the integral relation between
 $G^{\lambda}_{mn}$ and $G_{\cal P}$, can be regarded as the
 realization of the reduction procedure in the path integrals for the
 dynamical systems with a symmetry.

The reduction onto  the zero--momentum level, i.e., when $\lambda =0$,
establishes the  the relation between the path integrals that are
used for  descriptions of a quantum motion of the scalar particle on
an initial manifold 
$\cal P$   and on the the orbit space manifold $\cal M$.

In our case, in order to  represent the motion on the orbit
space, we have used an additional gauge surface $\xi _{\Sigma}$, on
which  the corresponding diffusion was given by the stochastic
differential equations (\ref{sd8}).  In these equation there is an
"extra" term, the drift  $j_{\Roman 2}$,  which is not directly
related to the orbit space $\cal M$. Without this term we would have
the stochastic process which could completely  correspond to the
diffusion on the orbit space. 

In path integrals the transformation in which we change the stochastic
process $\xi _{\Sigma}$ with the local stochastic differential
equation (\ref{sd8}) for the process $\tilde{\xi} _{\Sigma}$, with the
stochastic differential equation
\begin{equation}
dQ^{*}{}^{A}(t)={\mu}^2\kappa\biggl(-\frac12
G^{EM}N^C_EN^B_M\,{}^H{\Gamma}^A_{CB}+
j^{A}_{\Roman 1}\biggr)dt 
+\mu\sqrt{\kappa}
N^A_C\tilde{\mathfrak X}^C_{\bar M}dw^{\bar M},
\label{82}
\end{equation}
can be made with the help of the Girsanov transformation formula.

In our case, because of the  presence of  the projection operators in
diffusion matrices of equations (\ref{sd8}) and (\ref{82}) we have   
the degenerate diffusion matrices. 
It restricts the application of the standard Girsanov formula.
 
 But if we will remain in the frame of the predefined
  ambiguities, that originate from  using of the projection operators,
  we can still derive  the   Girsanov formula. In our case it will be
  also based on the uniqueness (modulo the above ambiguity) of the
  solution of the parabolic differential equation with the operator
  given by the diagonal part of the operator (\ref{op2}) and on the
  application of the It\^o differentiation formula for the composite
  function together with the account of  the following formula:
\[
\bigl(G^{AB}N^C_AN^D_B\bigr)\bigl(
(P_{\bot})^E_D
G^H_{EM}
(P_{\bot})^M_L\bigr)=(P_{\bot})^C_L\,.
\]

In result, the Radon--Nicodim derivative  of the measure
${\mu}^{{\xi}_{\Sigma}}$  with respect to the measure
${\mu}^{\tilde{\xi}_{\Sigma}}$ will be as follows: 
\begin{eqnarray*}
&&
\frac{d{\mu}^{{\xi}_{\Sigma}}}
{d{\mu}^{\tilde{\xi}_{\Sigma}}}
({\tilde{\xi}_{\Sigma}}(t))=
\exp\int^t_{t_a}\left[
-\frac12{\mu}^2\kappa \left((P_{\bot})^L_A
G^H_{LK}(P_{\bot})^K_E\right)
j^A_{\Roman 2}
j^E_{\Roman 2}dt
\right.
\nonumber\\
&&\left.
+\mu\sqrt{\kappa}G^H_{LK}(P_{\bot})^L_A
j^A_{\Roman 2}\tilde{\mathfrak X}^K_{\bar M}dw^{\bar M}
\right].
\end{eqnarray*}

Performing such a change of the integration variables in that  path
integral which is obtained as a result of the reduction to $\lambda
=0$ momentum level, we get the following integral relation: 
\[
G_{\Sigma}(Q^*_b,t_b;Q^*_a,t_a)=
\int_{\cal G}{G}_{\cal P}(p_b\theta ,t_b;
p_a,t_a)d\mu (\theta ),
\]
where the kernel $G_{\Sigma}$
 is presented by the path integral 
\begin{eqnarray*}
&&G_{\Sigma}(Q^*_b,t_b;Q^*_a,t_a)=\int_{ 
{\tilde{\xi}_{\Sigma}(t_a)=Q^*_a}\atop
{\tilde{\xi}_{\Sigma}(t_b)=Q^*_b}}
d\mu ^{\tilde{\xi}_{\Sigma}}\exp 
\left\{\frac 1{\mu
^2\kappa m}\int_{t_a}^{t_b}
V(\tilde{\xi}_{\Sigma}(u))du\right\}
\nonumber\\
&&\times
\exp\int^{t_b}_{t_a}\left\{
-\frac18{\mu}^2\kappa 
G^{AB}N^D_AN^L_B
\left[{\gamma}^{\alpha\beta}G_{DC}
  ({\tilde{\nabla}}_{K_{\alpha}}
  K_{\beta})^C\right]\right.
    \nonumber\\
  &&\left.\times
  \left[{\gamma}^{\mu\nu}G_{LE}
  ({\tilde{\nabla}}_{K_{\mu}}
  K_{\nu})^E\right]dt
    +\frac12\mu\sqrt{\kappa}
  N^D_P
  \left[{\gamma}^{\alpha\beta}G_{CD}
  ({\tilde{\nabla}}_{K_{\alpha}}
  K_{\beta})^C\right]
  \tilde{\mathfrak X}^P_{\bar M}dw^{\bar M}
\right\}\nonumber\\
&&(Q^*=\pi_{\Sigma} (p)).
\end{eqnarray*}

The semigroup determined by this path integral   acts in the space of
the scalar functions given on ${\Sigma}$.

Remarks, that there is a difference between the  formula obtained here
and an analogous formula from \cite{Stor2}. In the present formula 
the reduction Jacobian has an additional stochastic integral. It is
possible to get  rid of this stochastic integral with the help of 
 the corresponding  It\^o identity.   As it needs an additional
 investigation we  does not make this  transformation in the present
 paper.

\section{Conclusion}
From  our path integral transformation it follows that the path
integral measure is not invariant under the reduction (the formulas
(\ref{38}) and (\ref{39})). 

The obtained reduction Jacobian reveals an interesting geometrical
structure. Namely, it  is related with the mean curvature vector  of
the orbit over  the point belonging to the base space in the principal
fiber bundle.
After replacement of the variables in the path integral this mean
curvature together with the mean curvature of the orbit space adds to
the standard drift term of the stochastic differential equation
(\ref{sd8}).

We may suppose that the sum of two mean curvature comes from the mean
curvature of the manifold $\cal P$ provided that it is considered as
being embedded in some manifold with a bigger dimension. 

{\bf Acknowledgments.}
I thank to A.V.Razumov for the discussion of various geometrical
problems, V.O.Soloviev and V.I.Borodulin for valuable advises and
help.

\appendix
\section*{Appendix A}
\section*{Stochastic differential equation on a submanifold}
\setcounter{equation}{0}
\def\theequation{A.\arabic{equation}}
Let the   manifold $\cal M$ be embedded into the smooth (compact)
finite dimensional  Riemannian manifold with a metric $G_{AB}(Q)$.
We assume that this embedding is locally given by the equations
$Q^A=Q^A(x^i)$, where $\{Q^A\}$ is a coordinate system on the external
manifold and $\{x^i\}$ -- on $\cal M$. Then, on $\cal M$ we have an
induced metric: $h_{ij}(x)=Q^A_i(x)Q^B_j(x)G_{AB}(Q(x))$.

The stochastic process $\xi (t)=\{x^i(t)\}$, with the differential
generator $1/2\,{\triangle}_{\cal
M}$ (${\triangle}_{\cal M}$ is a Laplace---Beltrami operator on $\cal
M$) can be determined by the solution of the stochastic differential
equation with the following local representation: 
\begin{eqnarray}
&&dx^k(t)=-\frac12h^{ij}(x(t)){\Gamma}^k_{ij}(x(t))dt
+X^k_{\bar m}(x(t))dw^{\bar m}(t),\nonumber\\
&&({\sum}_{\bar m} 
X^k_{\bar m}X^l_{\bar m}=h^{kl}).
\label{ap1}
\end{eqnarray}

Now we will define the same stochastic process, but we will make use
the variables $Q^A$ that are related with the external manifold.
We assume that the stochastic differential equation which describes
the stochastic process on a submanifold can be written as
\begin{equation}
dQ^A(t)= a^Adt +{\tilde{\mathfrak X}}^A_{\bar
M}dw^{\bar M}(t),
\label{ap2}
\end{equation}
where $a^A$ and $ {\tilde{\mathfrak X}}^A_{\bar M}(t)$ are some (and
not defined yet) functions of $Q(t)$. 
Also, we require that at the initial moment of time the process
$Q^A(t)$ be on the  submanifold $\cal M$.

In order to find the explicit expressions for the coefficients of
equation (\ref{ap2}) we will apply the It\^o differentiation formula
to the 
function $Q^A=Q^A(x^i(t))$. As for the stochastic variables $x^i(t)$,
we will assume that they satisfy the equation (\ref{ap1}).

Then, comparing the result of such a differentiation with the
expression in the right--hand side of (\ref{ap2}), we find that the
coefficient  $a^{A}$ is equal to
\begin{equation}
a^A=-\frac12Q^A_i(x(t))h^{kl}(x(t))
{\Gamma}^i_{kl}(x(t))+
\frac12Q^A_{ij}(x(t))h^{ij}(x(t)).
\label{ap3}
\end{equation}
But 
\begin{eqnarray}
&&h^{kl}(x){\Gamma}^i_{kl}(x)=
G_{AB}(Q(x))
\left(
Q^A_{kl}(x)+
\right.
\nonumber\\
&&\,\,\,\,\,\,+\left.
{\Gamma}^A_{CD}(Q(x))
Q^C_{k}(x)Q^D_{l}(x)
\right)h^{im}(x)Q^B_{m}(x)h^{kl}(x)
\label{ap4}
\end{eqnarray}
(see, for example,\cite{Betounes}).
Taking this into account and using the projection onto the tangent
space to the manifold $\cal M$: 
\[
N^C_B(Q(x))=G_{BA}(Q(x))Q^A_i(x)h^{ij}(x)Q^C_j(x),
\]
we can transform (\ref{ap3}) to another form
\begin{equation}
a^A=-\frac12N^A_Ph^{ij}Q^C_iQ^D_j{\Gamma}^P_{CD}-
\frac12N^A_PQ^P_{kl}h^{kl}+\frac12 Q^A_{kl}h^{kl}.
\label{ap5}
\end{equation}

Since the components of the mean curvature vector of the submanifold
is given by
\begin{eqnarray*}
&&j^D=\frac12({\delta}^D_B-
N^D_B)h^{ij}\left[{\nabla}_{Q^P_i\frac{\partial}
{\partial Q^P}}\left(Q^L_j\frac{\partial}{\partial
Q^L}\right)\right]^B\nonumber\\
&&\,\,\,\,\,\,\,\,=\frac12h^{ij}(Q^A_iQ^B_j{\Gamma}^D_{AB}+Q^D_{ij}-
N^D_CQ^A_iQ^B_j{\Gamma}^C_{AB}-N^D_CQ^C_{ij}),
\end{eqnarray*}
we can rewrite (\ref{ap5})  as follows:
\begin{equation}
a^A(Q(x))=-\frac12G^{EM}(Q(x))N^C_E(Q(x))N^B_M(Q(x))
{\Gamma}^A_{CB}(Q(x))+j^A.
\label{ap6}
\end{equation}
where $j^A$ is, in fact, the function given on a submanifold, i.e.,
$j^A\equiv j^A(Q(x))$.
This follows from the fact that  the mean curvature can be also
defined without using an explicit coordinate expression (for example,
by the Weingarten map).

Before proceeding to the determination of the  diffusion coefficient
${\tilde{\mathfrak X}}^A_{\bar M}(t)$
 we notice, that the difussion coefficients of the equations
 (\ref{ap1}) and (\ref{ap2}) are  defined only up to the orthogonal
 transformations.

From the equality
\[
{\tilde{\mathfrak X}}^A_{\bar M}dw^{\bar
M}=Q^A_iX^i_{\bar m}dw^{\bar m}, 
\]
which can be  derived from (\ref{ap2}) in result of the application of
the It\^o differentiation formula, it follows that
\[
\sum _{\bar M}{\tilde{\mathfrak X}}^A_{\bar
M}{\tilde{\mathfrak X}}^B_{\bar M}=\sum _{\bar
m}Q^A_iX^i_{\bar m}Q^B_jX^j_{\bar
m}=h^{ij}Q^A_iQ^B_j=G^{CD}N^A_CN^B_D.
\]
These  equations define ${\tilde{\mathfrak X}}^A_{\bar M}$:
\[
{\tilde{\mathfrak X}}^A_{\bar M}=N^A_C{{\mathfrak X}}^C_{\bar
M}, \,\,\,\,\,\,(\sum _{\bar M}{{\mathfrak X}}^D_{\bar
M}{{\mathfrak X}}^C_{\bar M}=G^{CD}).
 \]

At last, redefining the coordinates $Q^A(x(t))$ of the stochastic
process for new coordinates $Q^A(t)$ (together with the requirement,
that at the initial moment of time a new process was also given on a
submanifold ), we get the following local stochastic differential
equation for  the components of the stochastic process on a
submanifold $\cal M$:
\begin{equation}
dQ^A(t)=\left(-\frac12G^{EM}N^C_EN^D_M{\Gamma}^A_{CD}+j^A
\right)dt+N^A_C{\mathfrak X}^C_{\bar M}dw^{\bar M},
\label{ap7}
\end{equation}
where all the functions in  right--hand side of this equation depend
on $Q^A(t)$.

\end{document}